\documentclass[seceq]{ptptex}

\usepackage{graphicx}
\usepackage{wrapft}

\notypesetlogo                       %comment in if to eliminate PTPTeX 
\title{%        %You can use \\ for explicit line-break
Periodicities in X-ray Binaries from Swift/BAT Observations
}

\author{%       %Use \scshape  for the family name
R.  \textsc{Corbet}$^{1}$,
C.  \textsc{Markwardt}$^{2}$,
L. \textsc{Barbier}$^{3}$, 
S. \textsc{Barthelmy}$^{3}$, 
J. \textsc{Cummings}$^{4}$, 
N. \textsc{Gehrels}$^{3}$, 
H. \textsc{Krimm}$^{1}$, 
D. \textsc{Palmer}$^{5}$, 
T. \textsc{Sakamoto}$^{4}$, 
G. \textsc{Sato}$^{6}$, \& 
J. \textsc{Tueller}$^{3}$,
on behalf of the Swift/BAT Survey Team
}

\inst{%     %Affiliation, neglected when [addenda] or [errata]
$^{1}$NASA GSFC/USRA,
$^{2}$NASA GSFC/UMD,
$^{3}$NASA GSFC,
$^{4}$NASA GSFC/NRC,
$^{5}$LANL,
$^{6}$NASA GSFC/ISAS
}

\abst{%         %this abstract is neglected when [addenda] or [errata]

The Burst Alert Telescope (BAT) on board Swift has accumulated
extensive light curves for 265 sources (not including GRBs) in the
energy range 14 to 200 keV. We present here a summary of searches for
periodic modulation in the flux from X-ray binaries. Our results
include: determination of the orbital periods of IGR J16418-4532 and
IGR J16320-4751; the disappearance of a previously known 9.6 day
period in 4U 2206+54; the detection of a 5 hour period in the
symbiotic X-ray binary 4U 1954+31, which might be the slowest neutron
star rotation period yet discovered; and the detection of flares in
the supergiant system 1E 1145.1-6141 which occur at both periastron
and apastron passage with nearly equal amplitude. We compare
techniques of weighting data points in power spectra and present a
method related to the semi-weighted mean which, unlike conventional
weighting, works well over a wide range of source brightness.

}

\begin{document}

\maketitle

\section{Introduction}
The Swift satellite was launched November 20, 2004.
It carries 3 instruments: 
the X-ray Telescope (XRT),
the Ultraviolet/Optical Telescope (UVOT),
and the Burst Alert Telescope (BAT).
Swift is primarily a gamma-ray burst mission.
After the BAT detects a GRB Swift slews rapidly to enable
UVOT and XRT observations.

The BAT consists of a coded aperture mask with 0.52m$^2$ CdZnTe detectors.
\cite{rf:Barthelmy05}
It has a wide field of view (1.4 sr half-coded) and
covers the  14 - 195 keV energy range.
The BAT typically observes 50\% - 80\% of sky each day.  Sky coverage
is ``random'' because of GRB followups.  Although
the primary BAT purpose is to detect
GRBs, it has also obtained light curves for 265 other sources.
We give here a brief summary of searches for periods in X-ray binaries
using BAT light curves.

\section{Weighting Power Spectra}
To search for periodic signals we calculate power spectra
of the light curves.  There is
large variability in the size of BAT error bars.  The errors depend on
the length of each observation and the
location of a source in the field of view. When
computing power spectra of data with non-uniform errors it can often
be
advantageous to weight data points by their errors.\cite{rf:Scargle89}  
This is analogous
to the weighted mean and we calculate the power spectrum of
$y_i/\sigma_i^2$. We term this ``standard weighting''.
However, if the scatter in data values is large compared to the
error bar sizes
weighting by error bars can be inappropriate.
We find that weighting bright
sources can actually reduce the sensitivity of the power spectrum
(Fig. 1).

\begin{figure}
%\centerline{\includegraphics[width=7 cm,angle=-90]{f1bw2.ps}}
\centerline{\includegraphics[width=7 cm,angle=-90]{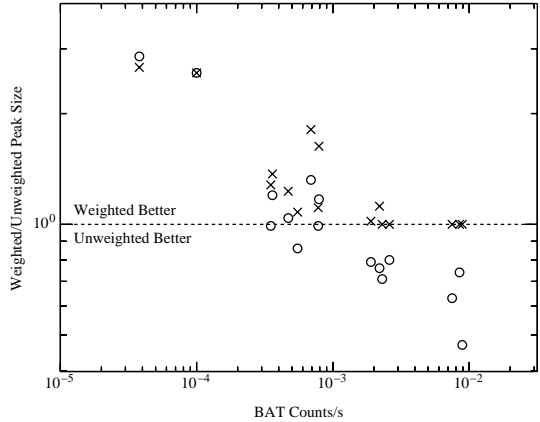}}
\caption{Comparison of sensitivity of weighting (circles) and
semi-weighting (crosses) power spectra.}
%\label{fig:1}
\end{figure}

A modified type of weighting is to 
treat source variability as an additional ``error'',
i.e. calculate the power spectrum of 
$ \frac{y_i}{((f \sigma_i)^2 + V_S)}$ where
$f$  is a correction to nominal error bar size and
$V_S$ is the estimated variance due to source variability.
Details are given in Corbet et al. (2007).\cite{rf:Corbet07}
This procedure is related to Cochran's semi-weighted
mean,
%\cite{rf:Cochran54}
\cite{rf:Cochran37,rf:Cochran54}
and hence may be termed ``semi-weighting''. We find
that semi-weighting works well for sources
across a wide range of brightness providing
that the correction factor $f$ is applied.
We advocate that semi-weighting should always
be considered when dealing
with data with non-uniform error bars.

\section{IGR J16418-4532 and IGR J16320-4751}

INTEGRAL has found many new highly absorbed high-mass X-ray binaries.
The BAT is well suited for studying these because of its high-energy
sensitivity.  
Two sources for which we have determined
orbital periods are IGR J16418-4532,
for which no pulsations have been seen, and 
the 1303s pulsar IGR J16320-4751.
We
find orbital periods of
P = 3.753 $\pm$ 0.004 days and
P = 8.96 $\pm$ 0.01 days respectively.\cite{rf:Corbet05,rf:Corbet06a}
For both of these sources, the strong absorption means that
the BAT is much more sensitive to orbital
modulation than the RXTE ASM which covers lower energies.
%This is despite the longer duration of the ASM light curves.

\begin{figure}[htb]
            \parbox{\halftext}{%   %\def\halftext{.471\textwidth}
 %               \figurebox{6cm}{2cm}
%   \centerline{\includegraphics[width=6.2 cm,angle=-90]{igr_bat_fts.ps}}
%   \centerline{\includegraphics[width=6.2 cm,angle=-90]{Corbet_Robin_fig2.ps}}
   \centerline{\includegraphics[width=6.2 cm,angle=-90]{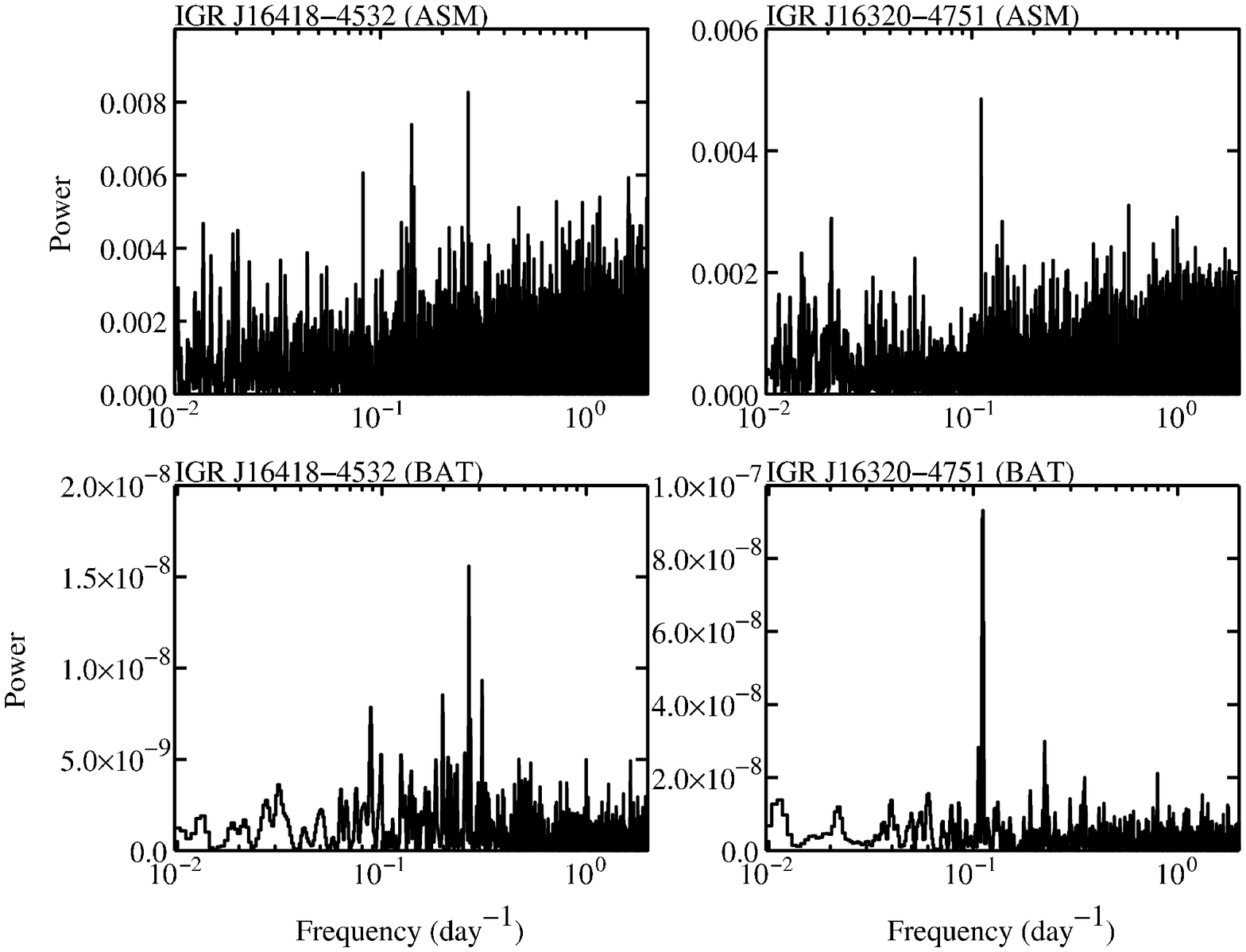}}
               \caption{Comparison of BAT (lower panels) and ASM 
(upper panels) power spectra of two INTEGRAL sources.}}
            \hfill
            \parbox{\halftext}{
%  \centerline{\includegraphics[width=6.2 cm,angle=-90]{1954_bat_ft.ps}}
%  \centerline{\includegraphics[width=6.2 cm,angle=-90]{Corbet_Robin_fig3.ps}}
  \centerline{\includegraphics[width=6.2 cm,angle=-90]{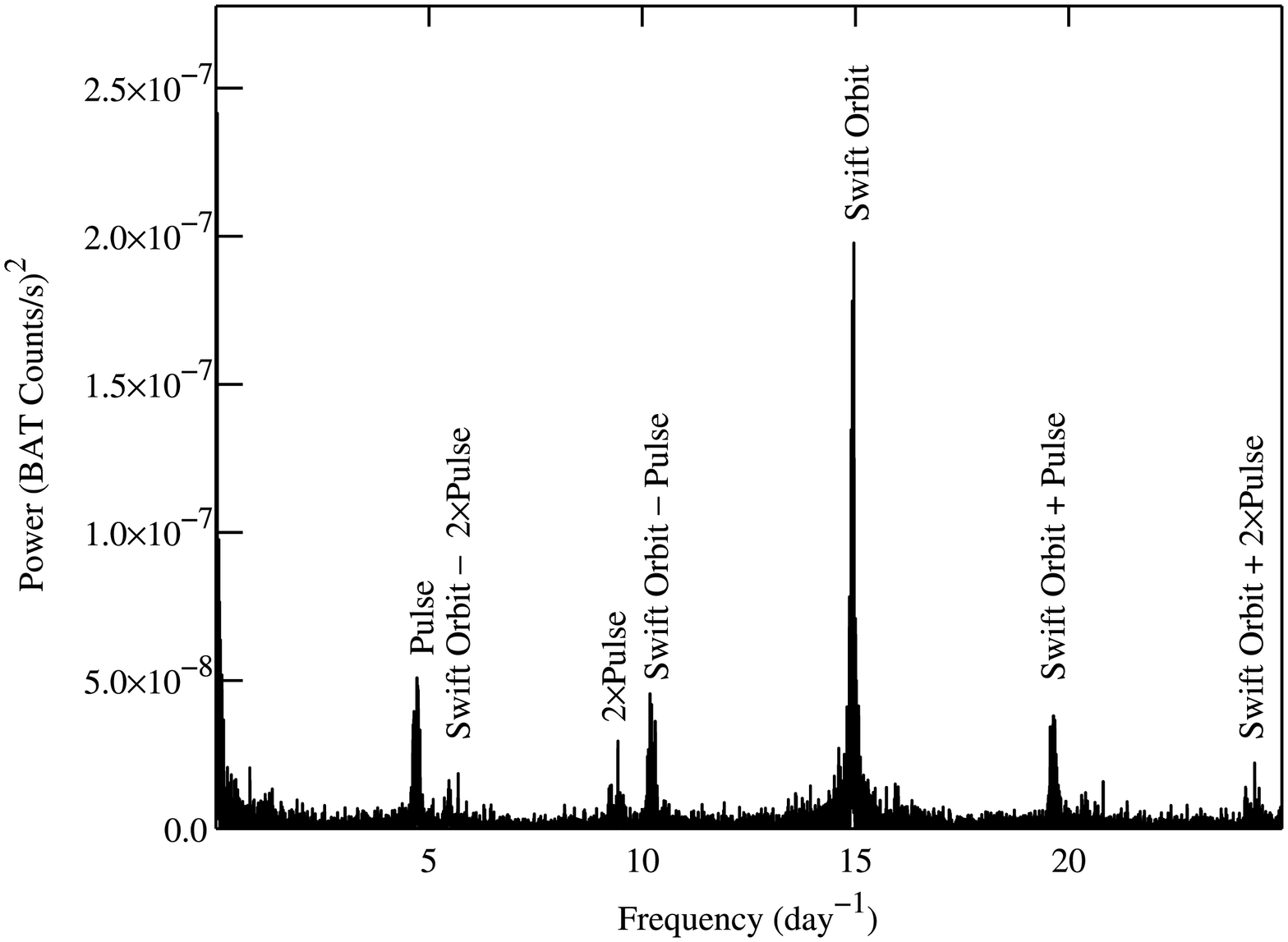}}
 %               \figurebox{6cm}{2cm}
                \caption{BAT power spectrum of 4U 1954+31. The
peaks are the pulse period, Swift orbital period, and interactions
between these.}}
\end{figure}

\section{An Ultra-Slow X-ray Pulsar in the Symbiotic System 
4U 1954+31}

Symbiotic binaries contain an object (typically a white dwarf)
accreting from an M giant.  4U 1954+31 was first thought to be a
high-mass X-ray binary, but an M
giant counterpart has now been found.\cite{rf:Masetti06}  
4U 1954+31 is thus a member of the
rare ``symbiotic X-ray binary''
class along with
GX 1+4 and 4U 1700+24.  4U 1954+31 is very variable, but no
orbital period or pulse period were previously known.

From BAT observations we discovered a strong 5 hour ``pulse''
period that decreased in period during an
outburst lasting hundreds of days.\cite{rf:Corbet06b}
This period is too short to be either
an orbital period or an M
star pulsation period.  The period change excludes
triple star models and also a white dwarf
rotation period.  The spin-up rate is consistent
with a neutron star pulse period if L$_X$ $\sim$ 5 $\times$ 10$^{35}$ ergs/s.  
This would be one of the slowest X-ray
pulsars known, currently only
exceeded by the 6.67 hr period source in RCW 103 for
which the origin is not yet clear.\cite{rf:DeLuca06}

\section{Disappearance of the 9.6 Day Period in 4U 2206+54}
4U2206+54 was previously thought to be a Be star X-ray pulsar.  
However, RXTE
observations (and a reanalysis of EXOSAT data) showed no pulsations.
Further optical observations suggest the optical
counterpart is a very peculiar
active O type star.  5.5 years of RXTE ASM data showed a 9.6 day
period.\cite{rf:CorbetPeele01}  If due to orbital variability, 
this would be one of
the shortest orbital periods known for a ``Be-like'' system.

The BAT data do not show modulation at 9.6 days.  Instead
the strongest peak in the
power spectrum is at 19.25 days, twice the ASM period.\cite{rf:Corbet07}  
Recent ASM
data folded on 19.25 days also show similar modulation
to the BAT light curve.  
The 9.6 day period is thus not a permanent strong feature of the
light curve and the orbital period 
thus may be twice the previously proposed
value. Possibly the circumstellar envelope and orbit
may not lie in the same
plane.

\section{The Supergiant High-Mass X-ray Binary 
1E 1145.1-6141}

1E 1145.1-6141 is an
X-ray pulsar with a 297 s pulse period.  
The optical counterpart is a B2
supergiant and accretion takes place from a stellar wind.  
The orbital period
was not determined until recently. Ray \&
Chakrabarty\cite{rf:Ray02} find from RXTE PCA pulse timing P = 14.365 $\pm$
0.002 days, and an eccentricity of 0.20 $\pm$ 0.03.

\begin{figure}[htb]
            \parbox{\halftext}{%   %\def\halftext{.471\textwidth}
 %               \figurebox{6cm}{2cm}
%   \centerline{\includegraphics[width=6.1 cm,angle=-90]{1145_bat_ft.ps}}
   \centerline{\includegraphics[width=6.0 cm,angle=-90]{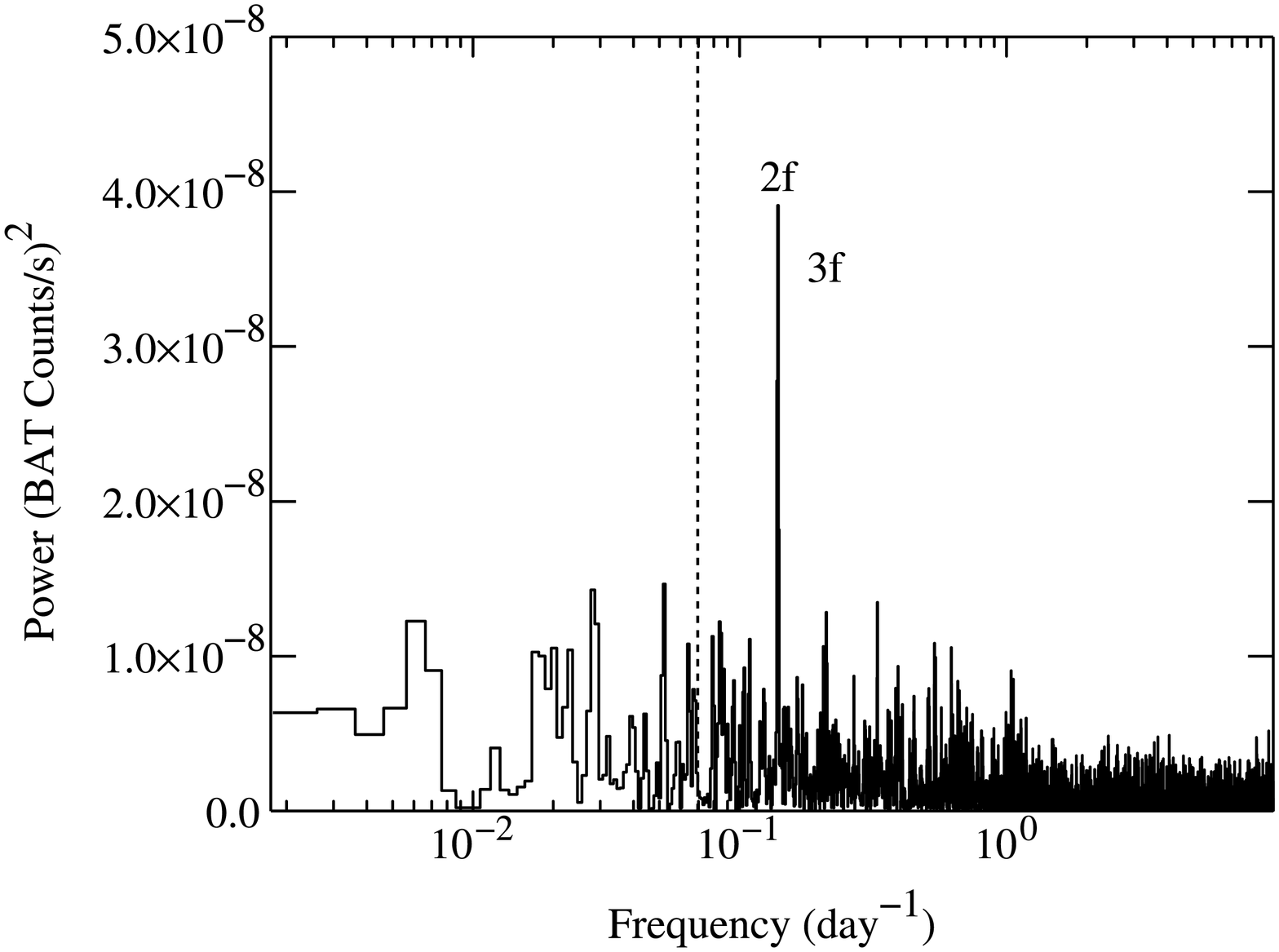}}
              \caption{Power spectrum of the BAT light curve
of 1E 1145.1-6141. The orbital period (dashed line) and harmonics are marked.}}
            \hfill
            \parbox{\halftext}{
%  \centerline{\includegraphics[width=6.1 cm]{fold_bat_asm.ps}}
  \centerline{\includegraphics[width=6.0 cm]{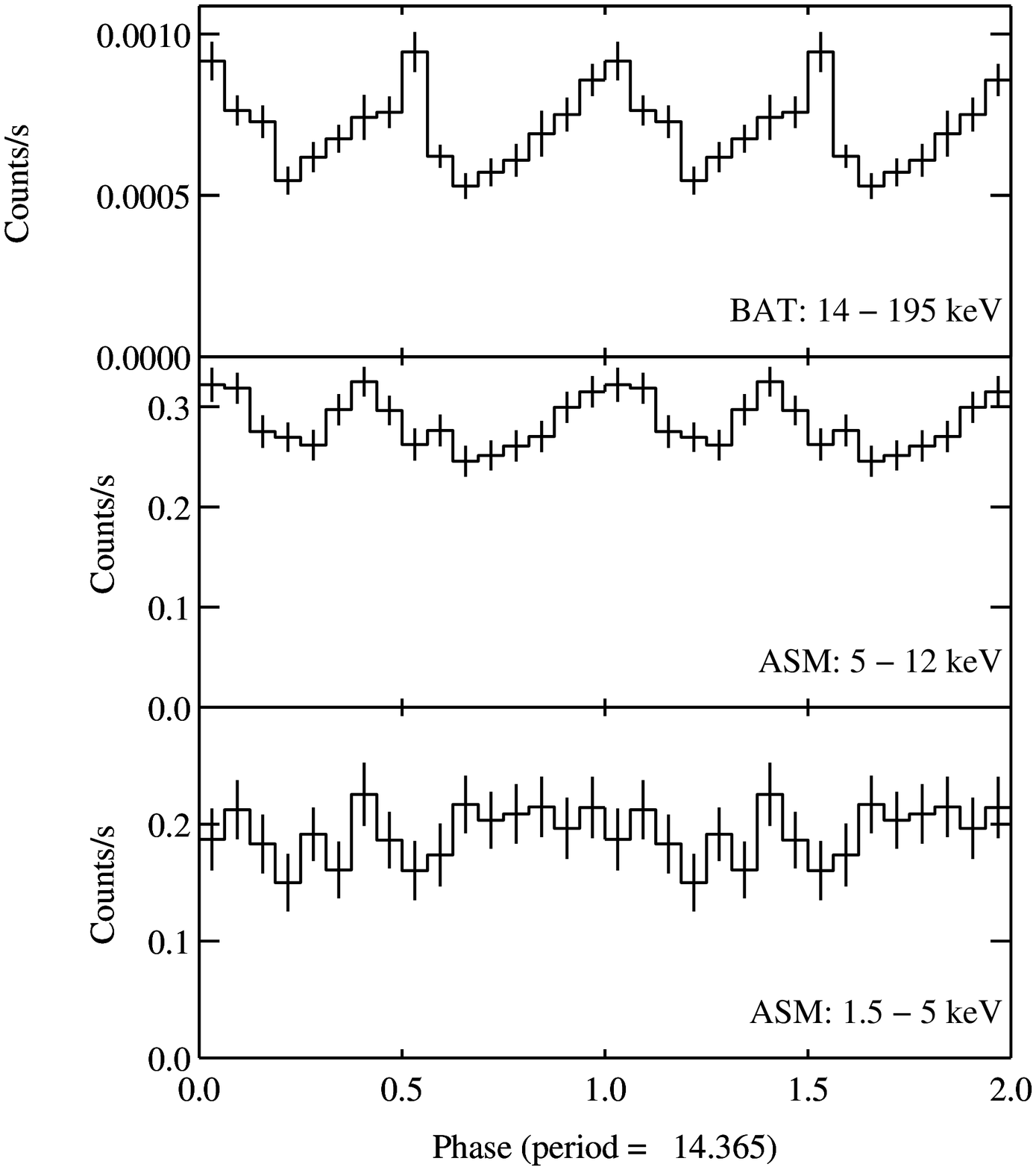}}
 %               \figurebox{6cm}{2cm}
                \caption{The BAT and ASM light curves of 1E 1145.1.6141
folded on the orbital period.}}
\end{figure}

1E 1145.1-6141 shows flares at apastron which  
are of similar size
to the periastron flares.  Apastron flares have also been reported from 
the supergiant systems 4U~1223-624\cite{rf:Pravdo01} 
and 4U~1907+09.\cite{rf:Roberts01}
However, 
in these two systems the apastron
flares are much smaller
than the periastron flares.
Apastron flare sources have low
inclination angles, suggesting that variability such as eclipses
in high inclination angle systems may mask apastron flares.
The origin of apastron flares is still unclear.

%\appendix
%\section{First Appendix} %Empty argument \section{} yields `Appendix'. 
%
%\section{Second Appendix}

\end{document}